\documentclass[preprint,aps]{revtex4}

\usepackage{graphicx}

\begin{document}

\title{Spectroscopic Evidence of Type II Weyl Semimetal State in WTe$_2$}
\author{Chenlu Wang$^{1,\sharp}$, Yan Zhang$^{1,\sharp}$, Jianwei Huang$^{1,\sharp}$, Simin Nie$^{1}$, Guodong Liu$^{1,*}$, Aiji Liang$^{1}$, Yuxiao Zhang$^{1}$, Bing Shen$^{1}$, Jing Liu$^{1}$, Cheng Hu$^{1}$, Ying Ding$^{1}$, Defa Liu$^{1}$, Yong Hu$^{1}$, Shaolong He$^{1}$, Lin Zhao$^{1}$, Li Yu$^{1}$, Jin Hu$^{2}$, Jiang Wei$^{2}$, Zhiqiang Mao$^{2}$, Youguo Shi$^{1}$, Xiaowen Jia$^{3}$, Fengfeng Zhang$^{4}$, Shenjin Zhang$^{4}$, Feng Yang$^{4}$, Zhimin Wang$^{4}$, Qinjun Peng$^{4}$, Hongming Weng$^{1,5}$, Xi Dai$^{1,5}$, Zhong Fang$^{1,5}$, Zuyan Xu$^{4}$, Chuangtian Chen$^{4}$ and X. J. Zhou$^{1,5,*}$}
\affiliation{
\\$^{1}$Beijing National Laboratory for Condensed Matter Physics, Institute of Physics, Chinese Academy of Sciences, Beijing 100190, China.
\\$^{2}$Department of Physics and Engineering Physics, Tulane University, New Orleans, Louisiana 70118, USA
\\$^{3}$Military Transportation University, Tianjin 300161, China.
\\$^{4}$Technical Institute of Physics and Chemistry, Chinese Academy of Sciences, Beijing 100190, China.
\\$^{5}$Collaborative Innovation Center of Quantum Matter, Beijing 100871, China.
\\$^{\sharp}$These people contributed equally to the present work.
\\$^{*}$Corresponding author: gdliu\_arpes@iphy.ac.cn, XJZhou@iphy.ac.cn.
}
\date{April 14, 2016}

\maketitle

\newpage

%%{\bf Abstract}

{\bf Quantum topological materials, exemplified by topological insulators, three-dimensional Dirac semimetals and Weyl semimetals,  have attracted much attention recently because of their unique electronic structure and physical properties.  Very lately it is proposed that the three-dimensional Weyl semimetals can be further classified into two types. In the type I Weyl semimetals, a topologically protected linear crossing of two bands, i.e., a Weyl point, occurs at the Fermi level resulting in a point-like Fermi surface. In the type II Weyl semimetals, the Weyl point emerges from a contact of an electron and a hole pocket at the boundary resulting in a highly tilted Weyl cone. In type II Weyl semimetals, the Lorentz invariance is violated and a fundamentally new kind of Weyl Fermions is produced that leads to new physical properties. WTe$_2$ is interesting because it exhibits anomalously large magnetoresistance. It has ignited a new excitement because it is proposed to be the first candidate of realizing type II Weyl Fermions. Here we report our angle-resolved photoemission (ARPES) evidence on identifying the type II Weyl Fermion state in WTe$_2$. By utilizing our latest generation laser-based ARPES system with superior energy and momentum resolutions, we have revealed a full picture on the electronic structure of WTe$_2$. Clear surface state has been identified and its connection with the bulk electronic states in the momentum and energy space shows a good agreement with the calculated band structures with the type II Weyl states. Our results provide spectroscopic evidence on the observation of type II Weyl states in WTe$_2$. It has laid a foundation for further exploration of novel phenomena and physical properties in the type II Weyl semimetals.}

The study of quantum topological materials has seen a rapid progress from the discovery of topological insulators and superconductors\cite{HasanReveiw,QiReview}, to three-dimensional Dirac semimetals\cite{Z.J.Wang1,Z.K.Liu1,S.Y.Xu1,Z.J.Wang2,Z.K.Liu2,M.Neupane,S.Borisenko,H.M.Yi} and to three-dimensional Weyl semitemtals\cite{H.Weyl,H.B.Nielsen1,S.Murakami1,X.G.Wan,A.A.Burkov,G.Xu,S.M.Huang,H.M.Weng,S.Y.Xu2,B.Q.Lv,L.X.Yang}. These materials have attracted much attention because they represent new states of matter with unique electronic structure, spin texture and associated novel physical properties. The latest development is the further classification\cite{A.Soluyanov} of the three-dimensional Weyl semimetals which resulted in the so-called "type II" Weyl semimetal in addition to the known type I weyl semimetals\cite{H.Weyl,H.B.Nielsen1,S.Murakami1,X.G.Wan,A.A.Burkov,G.Xu,S.M.Huang,H.M.Weng,S.Y.Xu2,B.Q.Lv,L.X.Yang}. In the type I Weyl semimetals, a topologically protected linear crossing of two bands, i.e., a Weyl point, occurs at the Fermi level resulting in a point-like Fermi surface. In this new type II Weyl semimetals, the Weyl point is expected to emerge from a contact of an electron and a hole pocket at the boundary resulting in a highly tilted Weyl cone. In type II Weyl semimetals, the Lorentz invariance is violated and a fundamentally new kind of Weyl Fermions is produced that leads to new physical properties\cite{A.Soluyanov}.

WTe$_2$ became well-known for its manifestation of extremely large magnetoresistance that was attributed to the compensation of electrons and holes in the material\cite{1AMN_N_2004}. Various alternative mechanisms have been proposed later on for accounting for the anomalous transport properties of WTe$_2$\cite{13PXC_ARXIV_2015,14JJ_PRL_2015,15LYK_APL_2015,16RD_PRB_2015,17WYL_ARXIV_2015}. A complete understanding of the electronic structure of WTe$_2$ is a prerequisite for ascertaining the origin of its anomalous transport properties. However, because of the existence of multiple pockets in a limited momentum space, the full nature of the electronic structure in WTe$_2$ remains controversial\cite{12PI_PRL_2014,8ZZW_PRL_2015,9ZYF_PRL_2015,10XFX_EPL_2015,11CPL_PRL_2015}. Lately, WTe$_2$ has ignited another surge of excitement because it is theoretically predicted to be the first material candidate that may realize ¡°type-II¡± Weyl state\cite{A.Soluyanov}. It has motivated exploration of type II semimetal state in other candidates like MoTe$_2$ and (Mo,W)Te$_2$ which are isostructural with WTe$_2$\cite{A.Soluyanov,Y.Sun,T.R.Chang,Z.J.Wang3,I.Belopolski}. However, so far no experimental evidence have been reported on the identification of Type II Weyl Fermions in WTe$_2$.

In this work, we report a combined experimental and theoretical study on the complete picture of the electronic structure and the existence of type II Weyl Fermions in WTe$_2$.  Taking advantage of our latest generation laser-ARPES system that can cover two-dimensional momentum space simultaneously with unprecedented energy and momentum resolution, we have revealed a full picture on the electronic structure of WTe$_2$.  We have unambiguously identified the existence of surface state that connects the bulk electron and hole pockets. High temperature ARPES measurements make us possible to reveal electronic states above the Fermi level.  The observed connection of the surface state with the bulk bands, its momentum evolution, its momentum and energy locations, are all in good agreement with the calculated band structures with type II Weyl points. Our results provide strong evidence on WTe$_2$ being a type II Weyl semimetal and lay a foundation for further exploration of novel physical properties in type II Weyl semimetals.

WTe$_2$ is a layered compound which crystallizes in the Td type of orthorhombic crystal structure\cite{45BBE_AC_1966}. Its crystal structure is constructed from layer-stacking along the {\it c}-axis of the two-dimensional WTe$_2$ sheets (as seen in Fig. 1a) (See Methods for details of the samples). Its corresponding bulk Brillouin zone and projected (001) surface Brillouin zone are shown in Fig. 1b and 1c, respectively. The ARPES data on WTe$_2$ are taken using our new laser-based ARPES system equipped with the latest-generation time-of-flight analyzer(see Methods for experimental details). It not only has super-high energy and momentum resolutions, but also has a unique capability of covering two-dimensional momentum space simultaneously, making it ideal in resolving fine electronic structure in the momentum space. We also take ARPES measurements under two distinct {\it s}- and {\it p}-polarization geometries. Because photoemission involved matrix element effects that enhance or suppress the signal of each bands, our measurements under two polarization geometries provide complementary information to fully reveal the electronic structures of the sample that is measured.

Depending on different cleavage surfaces and different measurement conditions, we have resolved three forms of distinct electronic structures on WTe$_2$ (see Fig. S1 in Supplementary Materials). The one reported in this main text is observed most often during the experiments, and this is also the same electronic structure form where the type II Weyl semimetal state is predicted in WTe$_2$. Therefore, we will focus on this electronic structure form in the main paper.

Figure 1(d,e) shows the Fermi surface of WTe$_2$ at 20 K measured under two different polarization geometries. The corresponding constant energy contours are shown in Fig. 1(i,j). For comparison, the calculated bulk Fermi surface, and the Fermi surface including the surface state (see Methods for band structure calculations), are shown in Fig. 1g and Fig. 1h, respectively. The band structures along some representative momentum cuts are shown in Fig. 2, and around some particular momentum space are shown in Fig. 3. It is clear that our super-high resolution laser ARPES measurements provide high quality data with unprecedented clarity that helps in resolving fine structures and providing a full picture of the electronic structure of WTe$_2$.

Careful analyses of the measured ARPES results (Figs. 1, 2 and 3), compared with band structure calculations, provide a measured Fermi surface picture of WTe$_2$, as schematically shown in Fig. 1f. The observed electronic structures can be summarized as the following.  (1). Three hole pockets are resolved, labeled as $\alpha$, $\beta$ and $\gamma$ in Fig. 1f, with their corresponding bands shown in Fig. 2f. (2). Two electron pockets can be resolved as $\delta$ and $\epsilon$ in Fig. 1f and their corresponding band marked in Fig. 2f. These two electron-like $\delta$ and $\epsilon$ bands are nearly degenerate, but they are still distinguishable, as seen in  Fig. S2 in Supplementary Materials. With increasing binding energy, the hole pockets ($\alpha$, $\beta$ and $\gamma$) expand while the electron pockets ($\delta$ and $\epsilon$) shrink, as seen from Fig. 1i and 1j, consistent with their assignments. (3). There is spectral weight near $\Gamma$ point in the measurement under {\it s}-polarization geometry (Fig. 1e) that is strongly suppressed in the measurement under {\it p}-polarization geometry (Fig. 1d). A flat band can be observed near $\Gamma$ point in Fig. 2l. Detailed analysis indicates that this band top is about 5 meV below the Fermi level, therefore does not form a Fermi surface sheet. With increasing binding energy, it shows up as a hole pockets and grows in its area, as seen in Fig. 1i and 1j. (4). A prominent feature in the measured Fermi surface is the V-shaped SS1 segment that connects the hole pockets and electrons in Fig. 1d and 1e. It keeps in contact with the electron pocket with increasing binding energy but gradually detaches from the hole pockets (Fig. 1i an 1j). This feature is not present in the calculated bulk Fermi surface (Fig. 1g) and bulk band structure (Fig. 2m) but is consistent with the band structure calculations including the surface state (Fig. 1h and Fig. 2n). The remarkable agreement between the measured and calculated Fermi surface (Fig. 1(d,e) and Fig. 1h), as well as between the measured and calculated band structures (Fig. 2f and 2l and Fig. 2n), provides compelling evidence that this SS1 feature is the surface state expected from band structure calculations. (5). In the calculated band structures (Fig. 2n), there is another surface state band ss2. But this band is clear above the Fermi level and merges into bulk bands below the Fermi level. Therefore,this ss2 surface band is not resolved clearly in our measured band structures (Fig. 1f and 1l).

The clear resolving of all the electronic structures by our super-high resolution laser ARPES provides a full picture of the electronic structure of WTe$_2$ which helps resolve the controversies in the literature due to missing components and mis-assignment. It is clear that, within the similar energy range, we have resolved more complete bands than reported before in WTe$_2$\cite{12PI_PRL_2014,14JJ_PRL_2015,46WY_PRL_2015,48MA_PRB_2000}.  The excellent agreement between our experiments and band structure calculations provides an unambiguous identification of each band in WTe$_2$. The absence of the third hole pocket in the calculations (Fig. 1g) could be due to slight chemical potential difference; slight shifting down of the chemical potential can produce the third hole pocket in the calculation (see Supplementary Materials). In particular, the clear observation and assignment of the surface state SS1 is important for examining the electron-hole compensation picture for understanding the anomalously large magnetoresistance in WTe$_2$ because it was either not resolved or assigned incorrectly as a bulk band\cite{12PI_PRL_2014,14JJ_PRL_2015,46WY_PRL_2015,48MA_PRB_2000}. According to our results, we obtain hole concentration of 0.017${\AA}$$^{-2}$ ($\alpha$), 0.013${\AA}$$^{-2}$ ($\beta$) and 0.005${\AA}$$^{-2}$ (one $\gamma$ pocket) for three hole pockets and electron concentration of 0.009${\AA}$$^{-2}$ ($\delta$) and 0.008${\AA}$$^{-2}$ ($\epsilon$) for the two electron pockets. The overall hole concentration is apparently larger than that of the overall electron concentration, asking for the reexamination of the electron-hole compensation picture for understanding the anomalous transport properties in WTe$_2$. In addition, the flat band very close to the Fermi level near $\Gamma$ may also play important role in dictating the properties of WTe$_2$.

Our high-resolution and complete ARPES results also make it possible to examine on the topological nature of WTe$_2$, as predicted from band structure calculations.  The observation of the SS1 surface state is consistent with the theoretical expectation of the type II Weyl Fermions in WTe$_2$\cite{A.Soluyanov}. Fig. 3 shows detailed evolution of the band strucure with momentum covering covering the momentum space with the surface state SS, and the bulk electron pockets and hole pockets. Along $\Gamma$-X direction (Cut 1), the surface state band ss1 comes out of the electron band $\delta$ at a binding energy of $\sim$80 meV. It nearly overlaps with the hole band $\alpha$ at the Fermi level. With the momentum cuts moving away from the $\Gamma$-X direction, the surface state band ss1 and the hole band $\alpha$ separate at the Fermi level, while it moves up in energy, gradually approaches the electron band $\delta$, overlap and eventually disappears when the momentum cut moves away from the electron pocket window (Cut 6). This evolution can be best seen from Fig. 1i and 1j. At high binding energies, the surface state SS1 connects to the electron pocket $\delta$. With reducing binding energy, the V-shaped surface state bottom tip gradually approaches the hole pocket $\delta$ around the Fermi level. Therefore, the surface state SS1 connects the electron pocket and hole pocket, as expected from band structure calculations.

According to our band structure calculations, there are four pairs of type II Weyl points (W1,W2) in WTe$_2$ on the k$_z$ = 0 plane, as shown in Fig. 4c in the momentum space that is consistent with previous calculations\cite{A.Soluyanov}. The W1 Weyl point is located 56 meV above the Fermi level while the other W2 Weyl point is 89 meV above the Fermi level (Fig. 4f). Two surface state bands are expected, ss1 and ss2, that are non-trivial,as also stated before\cite{A.Soluyanov,Y.Sun,T.R.Chang}. Along the $\Gamma$-X direction (Fig. 4b, Cut 1), the surface state band ss1 comes out of the electron band $\delta$ and merges into the hole band $\alpha$. When the momentum cuts moves away from the $\Gamma$-X direction (Fig. 4b, Cut 2), the electron pocket and hole pocket touches at the first Weyl point W1 and the surface state joins the W1 Weyl point. After that (Fig. 4b, Cut 3), the surface state ss1 comes from the electron band and goes band to the electron band. But another surface state ss2 connects the hole band and also electron band at even higher energy. Further increase of k$_y$ (Fig. 4b, Cut 4) results in the second touching of electron and hole pockets at the Weyl point W2 where the surface state ss2 merge into the bulk band. Eventually (Fig. 4b, Cut 5), the electron band and hole band separate again, and the surface state ss1 comes from the electron band and goes back to electron band. These detailed information provide rich information  for experiments to compare and check on the Weyl points.

In order to reveal the Weyl points that lie above the Fermi level, we carried out high resolution ARPES measurements on WTe$_2$ at high temperatures, which makes it possible to observe electronic states above the Fermi level due to thermal excitations. Fig. 4a show measured band structures at 165 K along $\Gamma$-X direction with different k$_y$ that covers the momentum and energy window of the Weyl points. This make it possible to observe the top of the hole bands (Fig. 4a) which are similar to those in band calculations (Fig. 4b).  Although the signal is relatively weak in this case simply because it is from thermal excitations, and the data are not clear enough to resolve individual Weyl points, the overall agreement between the measurements (Fig. 4a) and calculations (Fig. 4b) is rather good. In particular, it is clear that along $\Gamma$-X (Fig. 4a, Cut 1), the surface state ss1 comes out from the electron band and merges into the hole band. But when it comes to Cut 5 (Fig. 4a), the surface state ss1 comes out from the electron band and merges back to electron band, showing an excellent agreement with theoretical calculations.

In summary, by taking comprehensive super-high-resolution ARPES measurements and performing band structure calculations, we have provided a complete picture of the electronic structure for WTe$_2$ that are consistent with band structure calculations. We have clearly identified a surface state that connects the electron pocket and hole pocket. Above the Fermi level and in the momentum space where the type II Weyl Points are expected, our experimental results show good agreement with the band structure calculations on the relation between the surface state and bulk band states. These observations, and their excellent agreement with theoretical calculations, provide clear evidence on the formation of type II Weyl Fermions in WTe$_2$.  Our work paves the way for uncovering the origin of anomalous transport properties and exploration of novel physical phenomena associated with the formation of type II Weyl Fermions in WTe$_2$.

\vspace{3mm}

{\bf Methods}

The WTe$_2$ single crystals were grown via two different technical approaches. One is the chemical vapour transport method  and the other is the self-flux method with Tellurium as the solvent. In the chemical vapor method, high quality WTe$_2$ single crystals were synthesized  with stoichiometric mixture of W/Te powder, using the chemical vapor transport with iodine as the transport agent. During the crystal growth, the temperature was set at 900 $^o$C and 800 $^o$C, respectively, for the hot and cold ends of the double zone tube furnace. The sheet-like single crystals with metal luster can be obtained at the cold zone after one week¡¯ growth.

The angle-resolved photoemission (ARPES) measurements were performed at our new laser-based ARPES system equipped with the 6.994 eV vacuum-ultra-violet (VUV) laser and the time-of-ight electron analyzer (ARToF10K by Scienta Omicron)\cite{36ZYX_ARXIV_2016}. This latest ARPES system is capable of measuring photoelectrons covering two-dimensional momentum space (kx, ky) simultaneously. Measurements were performed using both {\it s}- and {\it p}-polarization geometries with a laser repetition rate of 500 KHz. In the {\it s}-polarization geometry, the electric field vector of the incident laser is perpendicular to the plane formed by the incident light and the lens axis of the electron energy analyzer, while in the p-polarization geometry, the electric field vector of the incident light lies within such a plane.  The overall energy resolution was set at 1-5 meV, and the angular resolution was $\sim$$0.1^{\circ}$. All the samples were measured in ultrahigh vacuum with a base pressure better than 5$\times10$$^{-11}$ mbar. The crystal was cleaved in situ at 20 K.The Fermi level is referenced by measuring on a clean polycrystalline gold that is electrically connected to the sample.

The electronic structure calculations are performed by using the projector augment wave (PAW) method\cite{37BPE_PRB_1994,38KG_PRB_1999} as implemented in the Vienna ab initio simulation package (VASP)\cite{39KG_CMS_1996,40KG_PRB_1996}. The experimental lattice constants are employed in the electronic structure calculations. The exchange and correlation effects are treated using the generalized gradient approximation (GGA)\cite{41PJ_PRL_1996}. Spin-orbit-coupling is taken into account in the self-consistent iterations. The integration is done on a grid of k points with the size 13$\times$9$\times$4. The plane wave cut-off energy of 500 eV is used. The maximally localized Wannier functions (MLWFs) for the {\it d} orbitals of W atoms and {\it p} orbitals of Te have been constructed by using the WANNIER90 code\cite{42MN_PRB_1997,43SI_PRB_2001,44MA_CPC_2008}. The topological surface states have been calculated by using the Wannier functions.

\vspace{3mm}

%%$^{\sharp}$These people contributed equally to the present work.\\

%%$^{*}$Corresponding author: gdliu\_arpes@iphy.ac.cn, XJZhou@aphy.iphy.ac.cn.

\vspace{3mm}

\noindent {\bf Acknowledgement}\\
This work is supported by the National Science Foundation of China (11574367), the 973 project of the Ministry of Science and Technology of China (2013CB921700, 2013CB921904 and 2015CB921300) and the Strategic Priority Research Program (B) of the Chinese Academy of Sciences (Grant No. XDB07020300). The work at Tulane  is supported by the US Department of Energy under Grant DE-SC0014208.

\vspace{3mm}

\noindent {\bf Author Contributions}\\
C.L.W. ,Y.Z. and J.W.H contribute equally to this work. X.J.Z. and G.D.L. proposed and designed the research. J.H, J.W., Z.Q.M. and Y.G.S. contributed in sample growth. S.M.N., H.W.M., X.D. and Z.F. contributed in the band structure calculations. C.L.W., Y.Z., J.W.H., G.D.L., A.J.L., Y.X.Z., B.S., J.L., C.H., Y.D., D.F.L., Y.H., S.L.H., L.Z., L.Y., X.W.J., F.F.Z., S.J.Z., F.Y., Z.M.W., Q.J.P., Z.Y.X., C.T.C. and X.J.Z. contributed to the development and maintenance of the Laser-ARTOF system and related software development. C.L.W., Y.Z., G.D.L. and J.W.H. carried out the ARPES experiment. C.L.W., Y.Z., G.D.L. and X.J.Z. analyzed the data. G.D.L., C.L.W. and X.J.Z. wrote the paper with Y.Z., L.Y., A.J.L., S.M.N. and H.M.W.. All authors participated in discussion and comment on the paper.\\

\noindent{\bf Additional information}\\
Supplementary information is available in the online version of the paper.
%permissions information is available online at www.nature.com/reprints.
Correspondence and requests for materials should be addressed to G.D.L. or X.J.Z.

%%\vspace{3mm}

%%\noindent {\bf\large Additional information}\\
%%\noindent{\bf Competing financial interests:} The authors declare no competing financial interests.

\newpage

\begin{figure*}[tbp]
\begin{center}
\includegraphics[width=1.0\columnwidth,angle=0]{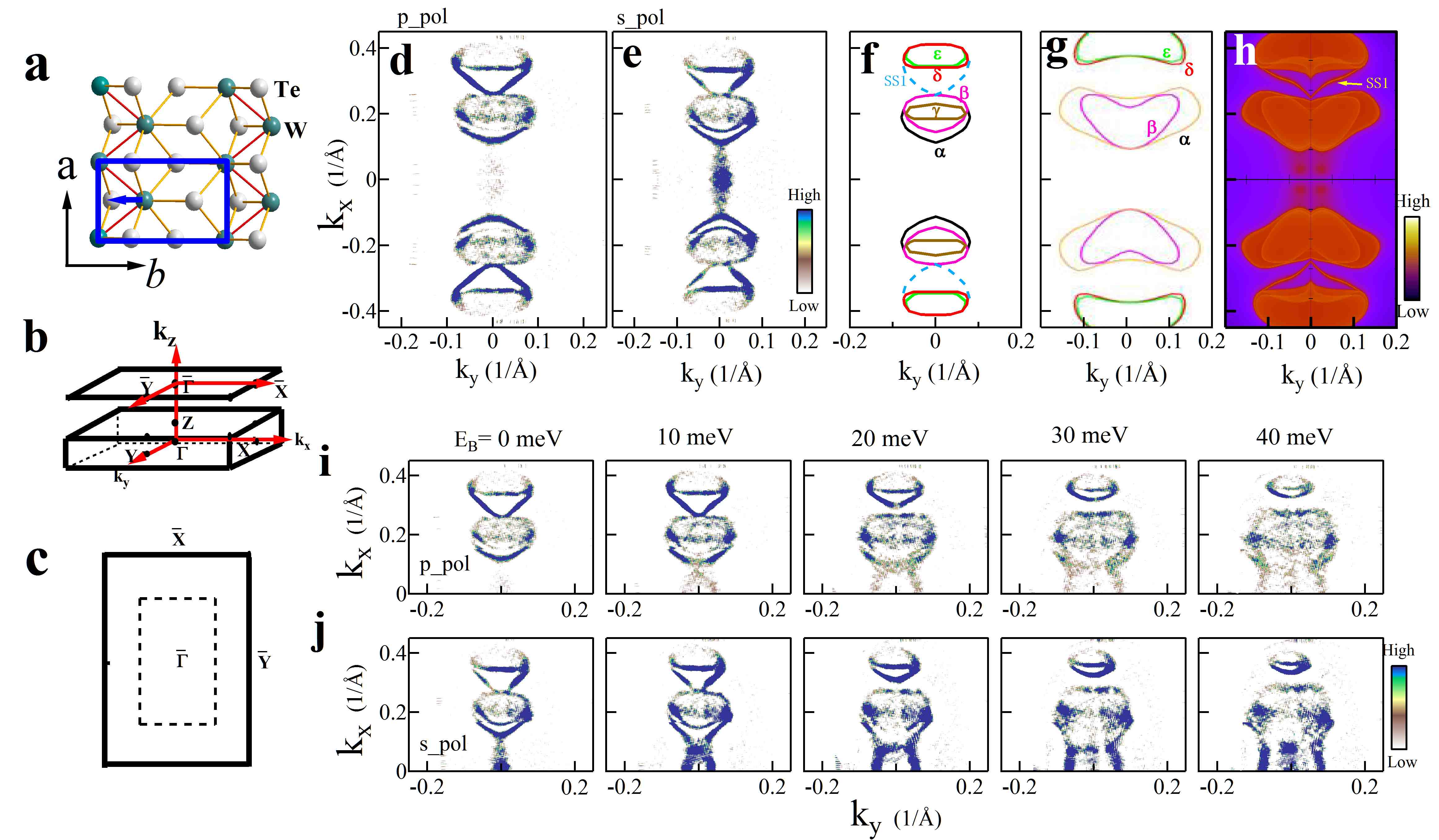}
\end{center}
\caption{{\bf Fermi surface of WTe$_2$ and its comparison with calculated results.} (a) Crystal structure of WTe$_2$ in Td form with a space group \emph{Pmn}2$_1$, viewed along the c axis (perpendicular to the stacked layers). The W-W zigzag chains are shown by the red solid line and the direction of the lattice distortion marked with the blue arrow. The unit cell is indicated by blue rectangle. (b) The bulk Brillouin zone and projected (001) surface Brillouin zone. (c) Corresponding (001) surface Brillouin zone. The dashed rectangle marks the momentum region covered in our ARPES measurements. (d) Fermi surface of WTe$_2$ taken by ARToF-ARPES at 20 K with {\it p}-polarization geometry. The spectral weight distribution is obtained by integrating photoemission spectra at each momentum within an energy window of [-2.5,2.5] meV with respect to the Fermi level. (e) Fermi surface of WTe$_2$ taken by ARToF-ARPES at 20 K with {\it s}-polarization geometry. In both (d) and (e), the lower half region is obtained by symmetrizing the upper half part of the measured Fermi surface with respect to the $\bar{\Gamma}$-$\bar{Y}$ mirror plane. (f) Schematic of all measured Fermi surface sheets. The solid lines represent the bulk pockets, and the dashed lines represent the surface state segments. Different pockets are drawn with different colors. (g) Calculated bulk Fermi surface at k$_z$=0. (h) Calculated projected Fermi surface including surface states. (i) Constant energy contours of WTe$_2$ at different binding energies measured at 20 K under {\it p}-polarization geometry. (j) Same as (e) but measured under {\it s}-polarization geometry. All the images in this figure are obtained by the second derivative of the original data with respect to energy.
}

\end{figure*}

\begin{figure*}[tbp]
\begin{center}
\includegraphics[width=1.0\columnwidth,angle=0]{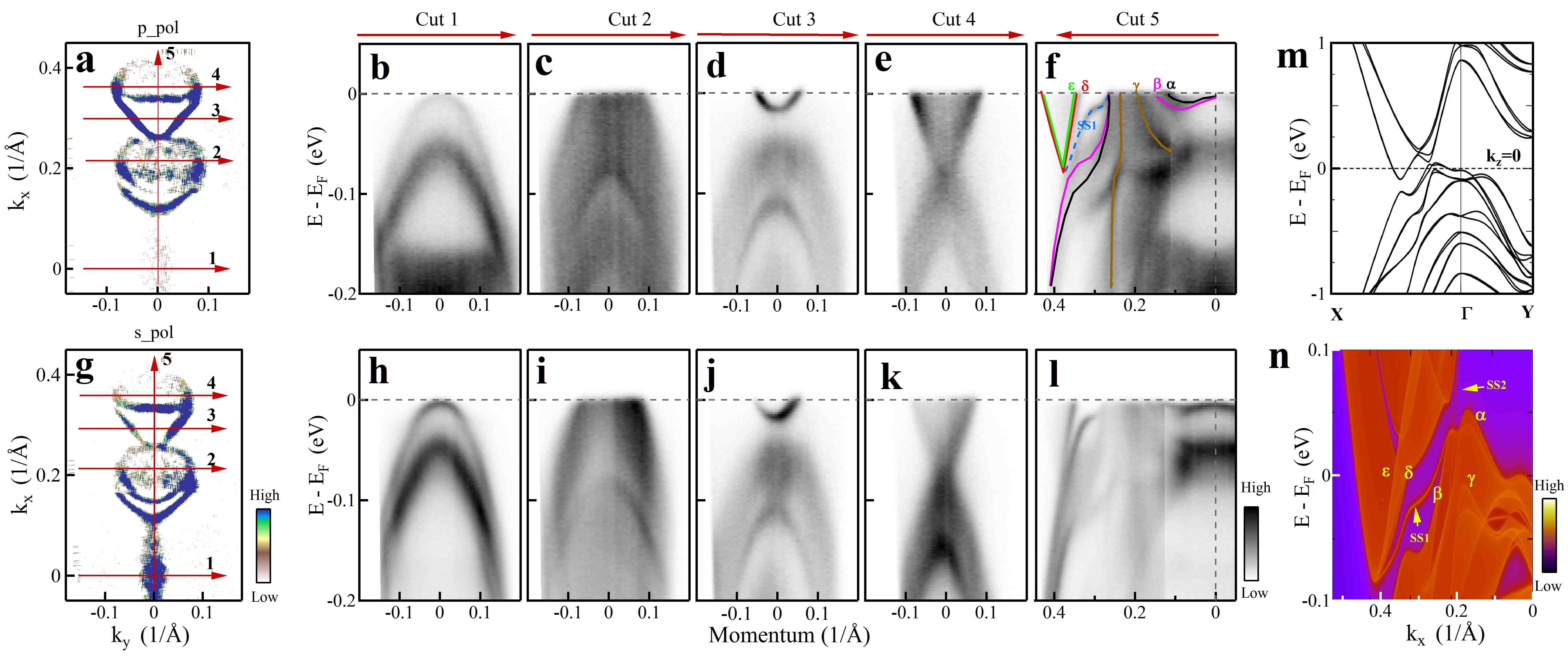}
\end{center}
\caption{{\bf Band structures measured along typical cuts of WTe$_2$ and their comparison with the band structure calculations.} (a) Fermi surface of WTe$_2$ measured at 20K under {\it p}-polarization geometry. The image is obtained by the second derivative of original data with respect to energy. (b$\sim$f) Band structures measured along different momentum cuts 1 to 5. The location of the five momentum cuts are marked by red lines in (a). The main bands are labeled in (f) by lines with their color corresponding to those of the Fermi pockets and surface state in Fig. 1f.  (g) Same as (a) but measured under {\it s}-polarization geometry. (h$\sim$l) Same as (b$\sim$f) but measured under {\it s}-polarization geometry. (m) Calculated bulk band structure along X-${\Gamma}$-Y with k$_z$ at 0. (n) Calculated band structure along ${\Gamma}$-X including surface states.
}

\end{figure*}

\begin{figure*}[tbp]
\begin{center}
\includegraphics[width=1.0\columnwidth,angle=0]{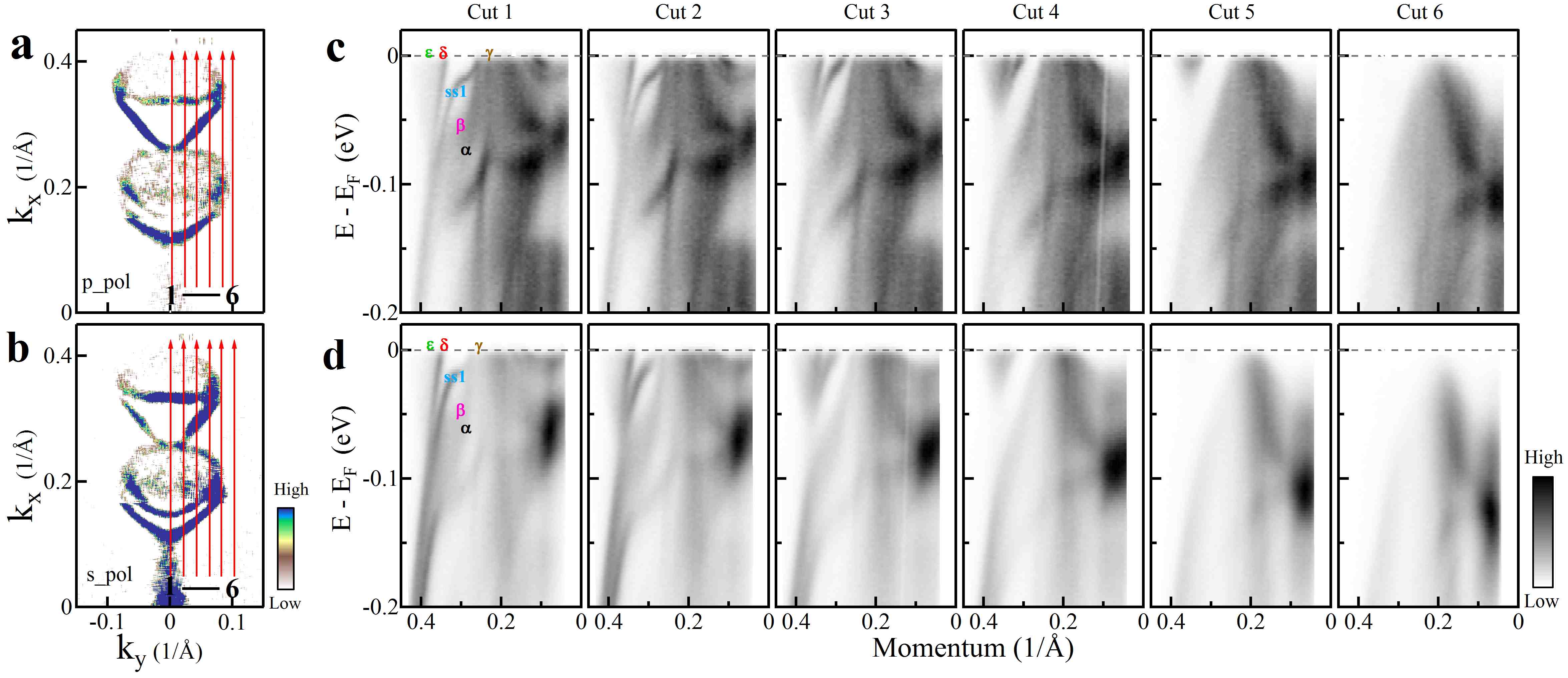}
\end{center}
\caption{{\bf Momentum evolution of the surface bands and the bulk bands.} (a) Fermi surface of WTe$_2$ measured at 20 K under {\it p}-polarization geometry. The image is obtained by the second derivative of original data with respect to energy. (b) Same as (a) but measured under {\it s}-polarization geometry.  (c) Band structures measured along the momentum cuts marked in (a) as red lines. They show clear momentum evolution of the surface bands and the bulk bands. (d) Same as (c) but measured under {\it s}-polarization geometry.
}

\end{figure*}

\begin{figure*}[tbp]
\begin{center}
\includegraphics[width=1.0\columnwidth,angle=0]{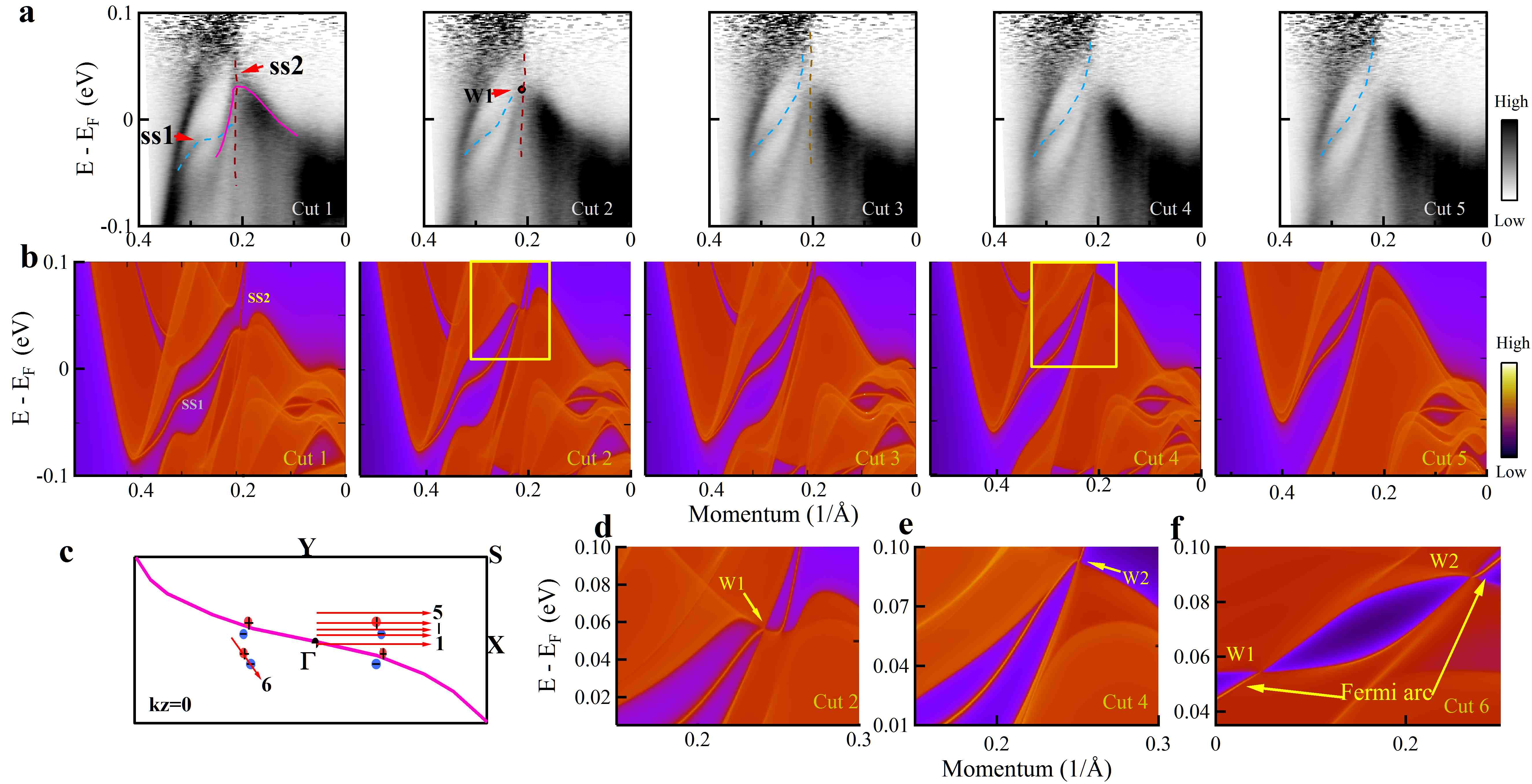}
\end{center}
\caption{{\bf Identification of Weyl points in WTe$_2$.} (a) Band structure of WTe$_2$ measured along different momentum cuts at 165 K. The location of the momentum cuts 1 to 5 are marked in (c). In order to observe electronic states above the Fermi level, the images are obtained by dividing the original data with the corresponding Fermi distribution function at 165 K. (b). The calculated band structures along the same five momentum cuts as marked in (c). (c). A schematic for the distribution of the calculated four sets of Weyl points within a bulk Brillouin zone at kz=0: Weyl point 1 (W1) represented by blue dots while Weyl point 2 (W2) by red dots.  The chirality is marked with "+" and "-", corresponding to C = +1 and C = -1. Wilson loop is shown schematically as the pink solid curve. The locations of the 5 momentum cuts are marked by red lines where Cut 2 crosses W1 Weyl point, Cut 4 crosses W2 Weyl point and Cut 6 crosses both W1 and W2 Weyl points. (d) Expanded view of the region marked by yellow rectangle in (b) for the Cut 2. Here the electron band and hole band touch at the W1 Weyl point. (e). Expanded view of the region marked by yellow rectangle in (b) for the Cut 4. Here the electron band and hole band touch at the W2 Weyl point. (f). Calculated band structure for the momentum cut 6 that crosses the two Weyl points.
}

\end{figure*}

\end{document}